\documentclass[aip,apl,reprint,superscriptaddress,floatfix]{revtex4-1}

\usepackage{graphicx}

\newcommand{\bs}{\begin{scriptsize}}
\newcommand{\es}{\end{scriptsize}}

\date{\today}

\begin{document}
\title{A tunable, dual mode field-effect or single electron transistor}
\author{B. Roche}
\affiliation{CEA, INAC, SPSMS, 17 rue des Martyrs, 38054 GRENOBLE Cedex 9, France}
\author{B. Voisin}
\affiliation{CEA, INAC, SPSMS, 17 rue des Martyrs, 38054 GRENOBLE Cedex 9, France}
\author{X. Jehl}
\email[]{xavier.jehl@cea.fr}
\affiliation{CEA, INAC, SPSMS, 17 rue des Martyrs, 38054 GRENOBLE Cedex 9, France}
\author{R. Wacquez}
\affiliation{CEA, INAC, SPSMS, 17 rue des Martyrs, 38054 GRENOBLE Cedex 9, France}
\affiliation{CEA, LETI, MINATEC Campus, 17 rue des Martyrs, 38054 GRENOBLE Cedex 9, France}
\author{M. Sanquer}
\affiliation{CEA, INAC, SPSMS, 17 rue des Martyrs, 38054 GRENOBLE Cedex 9, France}
\author{M. Vinet}
\affiliation{CEA, LETI, MINATEC Campus, 17 rue des Martyrs, 38054 GRENOBLE Cedex 9, France}
\author{V. Deshpande}
\affiliation{CEA, LETI, MINATEC Campus, 17 rue des Martyrs, 38054 GRENOBLE Cedex 9, France}
\author{B. Previtali}
\affiliation{CEA, LETI, MINATEC Campus, 17 rue des Martyrs, 38054 GRENOBLE Cedex 9, France}

\begin{abstract}
A dual mode device behaving either as a field-effect transistor or a single electron transistor (SET) has been fabricated using silicon-on-insulator metal oxide semiconductor technology. Depending on the back gate polarisation, an electron island is accumulated under the front gate of the device (SET regime), or a field-effect transistor is obtained by pinching off a bottom channel with a negative front gate voltage. The gradual transition between these two cases is observed. This dual function uses both vertical and horizontal tunable potential gradients in non-overlapped silicon-on-insulator channel.
\end{abstract}

\maketitle

\vspace{1cm}

Owing to its superior control of short channel effect together with a negligible dopant-induced variability \cite{Weber2008}, fully-depleted silicon-on-insulator (FD-SOI) is nowadays considered as a consistent solution for future low power applications \cite{Fenouillet-Beranger2009}.
One of the key challenge for FD-SOI is to design low access resistances. 
On the other hand single electron transistors (SETs) require access resistances of the order of the quantum unit (25.8\,$\mathrm{k\Omega}$) to exhibit Coulomb Blockade Oscillations (CBO)~\cite{Graber1991, Jehl2003}. Hence these two devices have so far been designed separately, although a simple modification of the Source/Drain architecture enables to get SET operation at low temperature with a metal-oxide-semiconductor field-effect transistor (MOS-FET) structure \cite{Hofheinz2006}.
Silicon SETs greatly benefit from the mature silicon technologies: recently scaling below the 5\,nm range allowed room temperature operation~\cite{Shin2010}. Coupled SET-FET circuits have been studied for multi-valued logic applications~\cite{Mahapatra2005}, realized with silicon technologies~\cite{Nishiguchi2004} or integrating CMOS devices~\cite{Prager2011}. Nevertheless a CMOS FD-SOI facility has never been used to realize such hybrid circuits, though it is an excellent tool to benchmark these concepts.
Here we report on the use of the substrate back gate to switch between FET and SET behaviour within the same device, fabricated in a CMOS facility.

\begin{figure}[!t]
 \centering
 \includegraphics[width=\linewidth,clip]{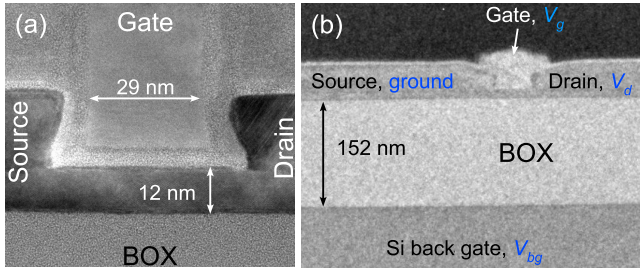}
 \caption{ a) Transmission electron microscope (TEM) image of a device showing the front gate (length $L_g=29$\,nm), channel (thickness $T_{Si}=12$\,nm) and buried oxide (BOX). The gate oxide thickness is 5\,nm. b) TEM image on a larger scale showing the various voltages applied during measurement, in particular the back gate substrate voltage ($V_{bg}$) applied across the BOX (150\,nm thick).}
 \label{fig:sample}
\end{figure}

We have fabricated n-MOSFETs adapted from FD-SOI technology (see Fig.\ref{fig:sample}). The SOI layer is etched to pattern the active area above the 150\,nm thick buried oxide (BOX). Silicon is then oxidised (5\,nm) and polycrystalline silicon is deposited, resulting in a conventional gate stack. After gate etching, the source and drain module is designed. For that purpose silicon nitride spacers are formed on both sides of the gate, epitaxy is performed to raise the source and drain, and finally arsenic is implanted, leading to a typical concentration above 10$^{20}$\,cm$^{-3}$. The resulting junction profile is such that the device is non-overlapped: the undoped region below the spacers---acting as a potential barrier for electrons---is responsible for the SET behaviour described hereafter~\cite{Hofheinz2006}. The silicon substrate below the BOX is used as a back gate.
This low-doped substrate used in industrial CMOS processes is not suitable for changing the voltage at low temperature: trying to change the back gate voltage leads to very slow relaxations, of the order of days, making experiments impossible. Shining light directly over the sample with a red LED thermally anchored at 4.2\,K and an optical fiber to transmit the light down to lower temperature stages, the substrate reacts much faster, making substrate polarisation studies possible. The experimental procedure we followed here consists in shining light during only a few seconds after each change of the back gate voltage value.

In this paper we show the data for two devices, however many samples have been produced and show the same behaviour. Fig.~\ref{fig:sample}a shows a transmission electron micrograph of a device similar to the samples we measured.

\begin{figure}[t]
 \centering
 \includegraphics{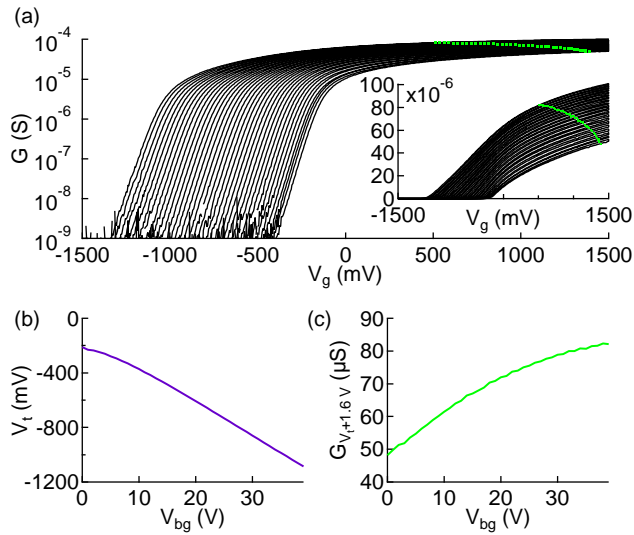}
 \caption{ a) Sample 1. Drain-source conductance $G$ versus front gate voltage $V_{g}$ at 300\,K for various $V_{bg}$, ranging from 0 (bottom) to 39\,V (top). b) Evolution of the threshold voltage $V_{t}$ with $V_{bg}$. The sub-threshold swing of 70\,mV/decade is nearly independent of $V_{bg}$. c) Drain-source conductance at $V_{g}=V_{t}$+1.6\,V, i.e. corrected for the shift of $V_{t}$. Note that it increases significantly with $V_{bg}$.}
 \label{fig:data1}
\end{figure}

The first sample has a channel thickness $T_{Si}=8$\,nm, an active width $W=40$\,nm and a gate length $L_g=70$\,nm. Considering the small $T_{Si}$ and the relatively long $L_g$ this sample is designed to have good sub-threshold electrostatic control by the gate. Its electrical characteristics at 300\,K are shown in Fig.~\ref{fig:data1}a in linear and logarithmic scales. It exhibits a sub-threshold slope of 70\,mV/decade (measured with $V_d=1$\,mV) which is at the state-of-the-art for an oxide thickness of 5\,nm, close to the theoretical limit for thermally activated transport $\frac{k_BT}{e}ln(10)$ (60\,mV/decade at 300\,K).

For such an non-overlapped geometry moderate on-current level is expected due to the extra access resistance to the channel. Fig.~\ref{fig:data1}c shows the variation of the conductance (in the linear regime) at $V_{g}=1.6$\,V above the threshold voltage. Using a positive back gate voltage $V_{bg}$, we increase by 20\,\% the normalized value of the on-current, reaching $0.4\,\mathrm{mA / \mu m}$ ($V_{bg}=39$\,V and the source-drain bias $V_{d}=1$\,V, not shown). This is a high value for an etched Si-nanowire.
Fig.~\ref{fig:data1}a shows that this gain is obtained without degrading the sub-threshold slope. Increasing $V_{bg}$ leads to a decrease of the threshold voltage ($V_t$), plotted in Fig.~\ref{fig:data1}b. From the slope $\mathrm{d}V_{bg} / \mathrm{d}V_t$ we can obtain the ratio of the effective front gate and back gate capacitances~\cite{Horiguchi2004}. We found a ratio of 40 which is not far from a crude estimation with planar capacitors: $\left| \mathrm{d}V_{bg} / \mathrm{d}V_t \right| \approx T_{BOX} / T_{ox} = 30$. The deviation from this model can be explained using a more realistic 3D model of capacitances. For practical applications much thinner BOX should be used to lower the applied $V_{bg}$.

\begin{figure}[t]
 \centering
 \includegraphics{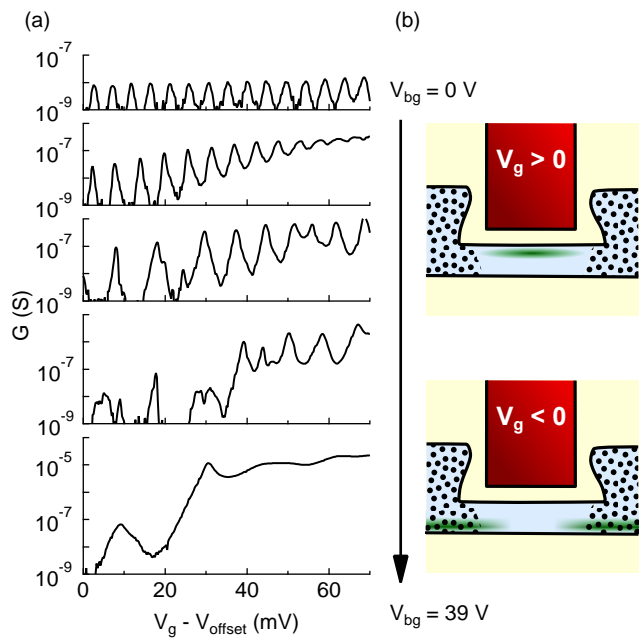}
 \caption{a) Sample 1. $G$ versus $V_{g}$ ($V_{d}= 100\,\mu V$) at 1\,K for $V_{bg} = 0, 12, 15, 18, 39\,V$ (from top to bottom). The curves have been shifted horizontally to show the onset of current ($V_{offset} = 400, -30, -210, -300, -890\,mV$ resp.). At $V_{bg}=0$\,V and $12$\,V regular CBO are observed. At $V_{bg}=39$\,V only very few aperiodic and broad features are observed attributed to residual single impurity effects. The crossover does not depend on the absolute value of the conductance $G$ but on $V_{bg}$ only. 
 b) Schematic picture of the device in the SET regime (top) and in the FET regime (bottom). The dotted areas correspond to high doping region acting as source drain reservoirs at low temperature. An electron gas (green) is either induced by the front gate (SET regime) or by the back gate (FET regime).
}
 \label{fig:cotun}
\end{figure}

At low temperature the non-overlapped geometry of our FET turns it into a SET without using the back gate, as reported before~\cite{Boeuf2004,Hofheinz2006}. Here we investigate how the device is modified by increasing the substrate voltage at $\mathrm{T}=1$\,K. Source-drain conductance is shown on Fig.~\ref{fig:cotun} for increasing $V_{bg}$. For each $V_{bg}$, the $V_{g}$ range is chosen to see the onset of current through the device. A clear SET regime is found at $V_{bg} = 0$\,V ; the observed CBO are regularly spaced with a period of $\Delta V_{g}=4.5$\,mV that corresponds to a gate capacitance $ C_{g}= {e \over \Delta V_{g}} \simeq 35$\,aF (consistent with the 27\,aF given by a planar capacitor model) and the value of conductance for each oscillation does not change significantly throughout the $V_{g}$ range. This SET regime is well described by the orthodox theory of Coulomb blockade because the island contains a large number of electrons at the onset of current. It is useful to note that the flat band condition is reached at $V_{bg} \simeq V_{g} \simeq 0$ in our devices at $\mathrm{T}=0$, we can therefor estimate that around one hundred electrons are present in the SET on Fig.~\ref{fig:cotun} at $V_{bg} = 0$. The situation changes gradually when increasing $V_{bg}$; the spacing between Coulomb peaks becomes larger and irregular, which characterizes a SET with a much lower density of electrons. The Coulomb island is progressively fragmented into low electron density flakes down to a Coulomb glass regime~\cite{Koulakov1997}. The electron island is well defined at $V_{bg} = 0$\,V because the electron gas experiences sharp gradient of potential at the top of the Si-wire, induced by $V_{g}$. On the contrary these gradients are smoother near the BOX interface, at a positive back gate voltages, leading to several electron flakes. Therefore the sharp transition between high electron density regions and potential barrier---which is necessary for orthodox SET---is progressively lost at large $V_{bg}$. Finally, for $V_{bg} = 39$\,V, an electron gas is located at the buried interface and pinched off by a negative front gate voltage, leading to a FET characteristic. The observed bumps below the threshold are due to remaining disorder affecting the smooth parabolic potential of the pinch-off region.

The second sample is designed with a reduced $L_g$ of 30\,nm. The slightly thicker $T_{Si}$ is now 12\,nm and W is unchanged. We aim to see the effect of scaling down $L_g$ both on the SET and the FET.

\begin{figure}[t]
 \centering
 \includegraphics[width=\linewidth,clip]{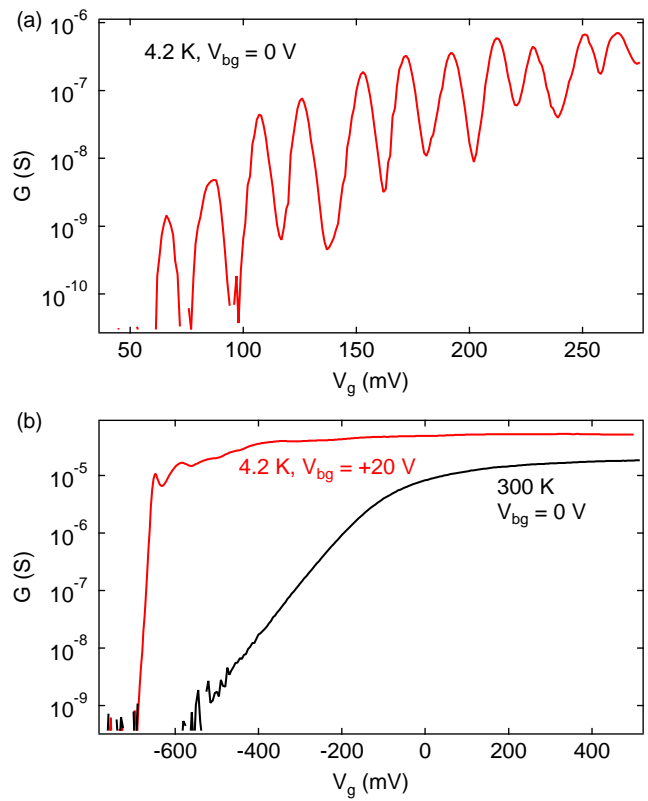}
 \caption{a) Sample 2. Coulomb blockade oscillations in the source-drain conductance G versus gate voltage at zero back gate voltage. b) In red, the characteristic of the same device at $V_{bg}$=+20\,V. The Coulomb oscillations are replaced by a typical field-effect curve showing an excellent sub-threshold slope of 8\,mV/decade. For comparison the characteristic at 300\,K and $V_{bg}$=0\,V is also shown.}
 \label{fig:set2fet}
\end{figure}

Fig.~\ref{fig:set2fet}a shows the characteristic of sample 2 at 4.2\,K and $V_{bg}=0$\,V. Regular CBO appear due to the non-overlapped geometry but the measured period $\Delta V_g=21$\,mV\,$\pm\,6$\,mV which corresponds to a capacitance $C_g = e / \Delta V_g \simeq 8$\,aF is larger by a factor 4 as compared to sample 1. This is mainly a consequence of the smaller gate length, that is the smaller gate-channel overlap capacitance.
In sample 2 the first few electrons in the island are detected (at $V_{g} \gtrsim 0$). This is due to the larger tunnel coupling to source and drain at larger $T_{Si}$. The low density limit explains the fluctuations of $\Delta V_{g}$ through quantum capacitance effects.

By applying $V_{bg}=+20$\,V (red line in Fig.~\ref{fig:set2fet}b) an excellent FET regime is now observed, with both a very steep current rise, and a good on-conductance. 
We have not significantly improved the on-current level as compared to sample 1 because the on-current level is limited by the access resistance.
The sub-threshold swing at $\mathrm{T}=4.2$\,K is excellent, reaching 8\,mV/decade. Compared to sample 1 less bumps of conductance are observed near the threshold, indicating that the pinch-off potential is steeper (smaller gate length) and less sensitive to the disordered potential.
The sub-threshold swing is nevertheless larger than the best theoretical limit at $\mathrm{T}=4.2$\,K which is 0.85\,mV/decade. We attribute the observed swing to the thermal activation with a lever arm parameter $\alpha = {\delta \phi / \delta V_{g}} \simeq 0.1$ ($ \phi$ is the potential at the electron gas location). At $V_{bg}=0$\,V from the analysis of the CBO we have measured $\alpha \simeq 0.3$ when the electron gas is located near the top interface. The lower $\alpha$ value found in the FET regime is consistent with the fact that the electron gas has been pushed towards the BOX interface.

For comparison the FET characteristic at 300\,K and $V_{bg}=0$\,V is shown on the same plot. It exhibits a sub-threshold swing of 112\,mV/decade, as a result of its short channel and relatively thicker $T_{Si}$. The maximum transconductance at 4.2\,K is $30\,\mu S$ (not shown), which makes this FET a good candidate to design an amplifier at cryogenic temperature\,\cite{Gremion2010}.

In conclusion, we report here the fabrication of a device acting either as a SET or a FET depending only on substrate voltage. The CMOS fabrication process enables large scale integration. We showed that scaling down the gate length both decreases the size of the SET down to the few electrons limit and improves the FET caracteristic at $\mathrm{T}=4.2$\,K.
This work opens up the opportunity to design hybrid circuits with SET and FET devices using a local back gate. Such circuits are well adapted to low power applications for which they will not suffer from the relatively high access resistance of the FETs due to the non-overlapped geometry.

The research leading to these results has received funding from the European Community's seventh Framework (FP7 2007/2013) under the Grant Agreement Nr:214989. The samples subject of this work have been designed and made by the AFSID Project Partners http://www.afsid.eu.

\end{document}